\documentclass[a4paper]{article}
\usepackage{graphicx}
\usepackage{amsmath}
\usepackage{hyperref}
\usepackage{color}

\begin{document}

\begin{center}

{\bf \Large To cooperate or to defect? Altruism and reputation}\\[5mm]

{\large  Krzysztof Ku{\l}akowski and Przemys{\l}aw Gawro\'nski}\\[3mm]

{\em

Faculty of Physics and Applied Computer Science,

AGH University of Science and Technology,

al. Mickiewicza 30, PL-30059 Krak\'ow, Poland

}

\bigskip

$^*${\tt kulakowski@novell.ftj.agh.edu.pl}

\bigskip

\today

\end{center}

\begin{abstract}
The basic problem in the cooperation theory is to justify the cooperation. Here we propose a new approach, where 
players are driven by their altruism to cooperate or not. The probability of cooperation depends also on the co-player's
reputation. We find that players with positive altruism cooperate and met cooperation. In this approach, payoffs are not 
relevant. The mechanism is most efficient in the fully connected network.
\end{abstract}





\section{Introduction and the model}

The Prisoner's Dilemma (PD) is perhaps the most famous paradox in the game theory \cite{pridi}. Each of two identical
players has two different strategies: to cooperate (C) with the other or to defect (D) the cooperation. The payoff
matrix is constructed as that for each player, the strategy D is always better than the strategy C. When both defect,
their payoff is worse than when both of them cooperate. The challenge is to understand the reason why people do
cooperate in situations, where PD applies. The spectrum if these situations is very rich: from the nuclear
disarmament to the environmental protection, from a competition between companies to disputes on child rearing.
However, to justify cooperation in the frames of PD seems to be impossible a priori. The minimal assumption of
the classical game theory is that players are rational, i. e. they do not play strategies where their payoff
is always worse. Such strategies are termed 'dominated'. In PD, the strategy of cooperation is dominated, {\it q.e.d.}\\

Faced with this difficulty, researchers tend to capture PD from more general point of view, as the evolutionary game
theory \cite{axel0}, where boundedly rational agents learn the best strategy in repeated games. These generalizations
met older discussions within sociology, to what extent the game theory can be applied in social sciences. Meanwhile 
the subject of the research changed from a herd of "undersocialized" actors, rationally looking for their individual benefits, 
towards a society of "oversocialized" citizens, obeying social norms which they have been internalized \cite{gran,raub}. 
Collective forms of cooperation entered the discussion, as the altruistic punishment \cite{fegach} or indirect 
reciprocity and reputation \cite{nosi}. Lists of references can be found, for example, in \cite{ax6,szafa,brandt,fefi}.
Most if not all of these references rely on the concept of payoffs. \\

Here we intend to formulate a minimal model of cooperation, without payoffs, and with reputation as a necessary ingredient.
Reputation - commonly accessible information on the individual scores - secures the collective character of the system 
dynamics. Scores are memorized as $W_i$. Each time when $i$ cooperates, $W_i$ is transformed as $W_i \to (W_i+1)/2$;
when $i$ defects, $W_i \to W_i/2$. This is a convenient way to memorize a few recent decisions of $i$. (Also in this way
we omit the complexity of subtle differences between cooperators with cooperators and defectors of defectors \cite{nosi}; 
we note that such a selectivity leads to a paradox, when a defector could qualify a defection of his own as the proper 
choice and an argument to cooperate.) Another element of our approach is the level of altruism $\epsilon_i$ of a player, as a measure 
of her/his willingness to cooperate with others or to defect. The probability that $i$ cooperates with $j$ is calculated as 

\begin{equation}
P(i,j)=W_j+\epsilon_i
\end{equation}
limited additionally to the range (0,1). Then, if $W_j+\epsilon _i>1$, $P(i,j)$ is set to 1. The altruisms $\epsilon _i$ are 
selected randomly from the uniform distribution in the range (-1/2,1/2) and they are not changed during the simulation. 

Looking for a similar approach in literature, we find that some our assumptions are close to those made by Levine \cite{levi}.
There, a player 1 calculates the utility of the co-player 2 and the one of himself with different weights. These weights depend
on the altruism of the player 1; also, the obtained utility depends on the altruism of the co-player 2. With these assumptions,
Levine analysed the data of the ultimatum bargaining and the public good experiments, and he got a reasonable accordance with 
these data. The model \cite{levi} was supposed to be a simpler version of some preceding approach \cite{rabin}. Our approach 
is even simpler in the sense that here we have no utility at all. The mutual interaction between players is parametrized as 
to reflect merely the influence of a particular decision on the probabilities of the other's decisions. In our former works, 
this shortcut was found to be quite efficient \cite{k1,k2}.

\section{Results}

In Fig. 1 we show the percentages of cases when the player $i$ was rewarded (R, both cooperated), defected when he 
cooperated (S), when $i$ defected when his co-player cooperated (T) and when both defected (U), against the altruism 
$\epsilon _i$. These results indicate that the altruism triggers a positive feedback: an increased probability to cooperate 
provides a good reputation what induces the willigness of others to cooperate. As the number of games increases, the curves
become more and more clear and thin; this is due to the fact that initial transient effects contribute less and less to the 
whole statistics. It is somewhat surprising that the same data are much less clear, when presented against the reputations 
$W_i$ (Fig.2). This is a consequence of strong fluctuations of $W$'s; in our model, a recent defection turns the reputation 
into ruins, i.e. to values less than 0.5.\\

\begin{figure}
\begin{center}
\includegraphics[scale=0.45,angle=-90]{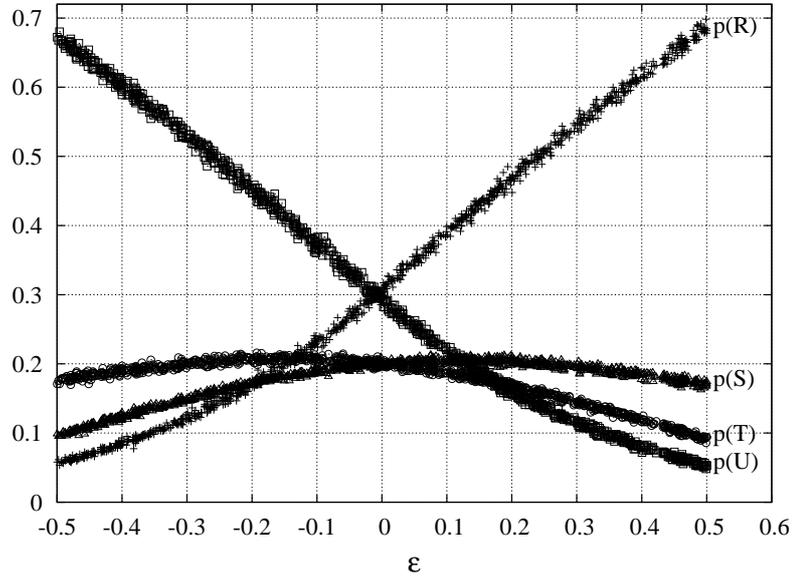}\\
\end{center}
\caption{Statistics of four different situations (R, S, T and U) against the altruism level $\epsilon$. The picture refers
to the state after $5\times 10^6$ games played in the set of 1000 players. For $\epsilon >0$, the mutual cooperation (R) is most 
likely. For $\epsilon <0$, the mutual defection (U) happens in most cases.
}
\end{figure}

\begin{figure}
\begin{center}
\includegraphics[scale=0.45,angle=-90]{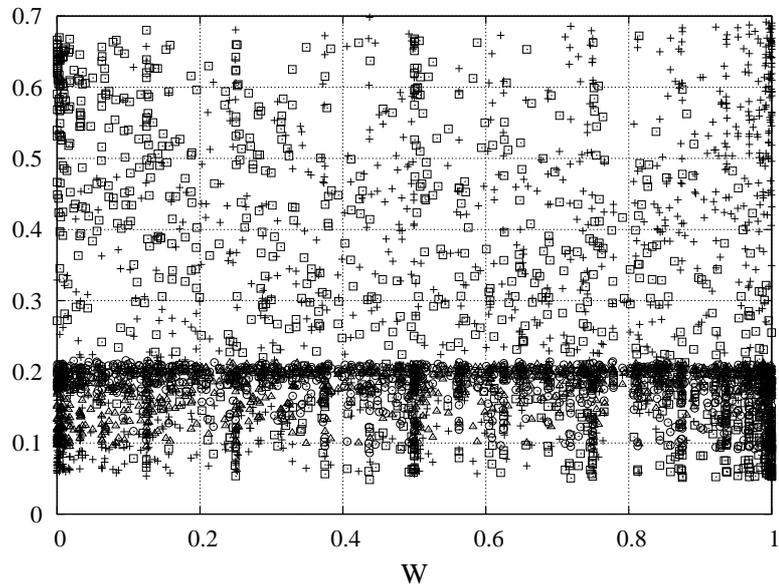}\\
\end{center}
\caption{Statistics of four different situations (R, S, T and U) against the reputation $W$. The picture refers to the state 
after $5\times 10^6$ games played in the set of 1000 players. Narrow black bands are produced by events, when a reputation 
close to 1 is reduced by factor $0.5$, $(0.5)^2$ etc.
}
\end{figure}

Trying to omit this difficulty by some kind of a mean field approach we can ask if there is a unique curve $W(\epsilon)$, 
where $0<W(\epsilon)<1$. Let us suppose that 
in stationary state, a player $i$ with infinitesimally small $\epsilon_i$ has equal chance to cooperate and to defect, and 
then his reputation does not change in the average. The probability that he cooperates is then equal to 1/2. For a given 
curve $W(\epsilon)$, this probability can be found as

\begin{equation}
P_i(Cooperate)=\int_{-1/2}^{1/2}d\epsilon_j [W(\epsilon_j)+\epsilon _i]=<W>+\epsilon_i
\end{equation}
where $<W>$ is the mean value of $W$. This equation has no solution, as $<W>=const(\epsilon_i)$. On the other hand, we can expect
that $W(-\epsilon)=1/2-W(\epsilon)$, and therefore $<W>$=1/2. The equation above can be fulfilled only for the altruism $\epsilon_i$=0.
Now let us consider the case when the function $W(\epsilon)=0$ for $\epsilon <0$, otherwise $W(\epsilon)=1$. Asking, if this 
function is stationary, consider a player with $\epsilon<0$. He
cooperates with probability 0 with those of $W=0$ and with probability $1+\epsilon<1$ with those of $W=1$. Then his mean 
probability of cooperation is $(0+1+\epsilon)/2<$, what is less than 0.5. Therefore his reputation decreases; however, according 
to the actual curve $W(\epsilon)$, it is already zero and cannot decrease any more. Now let us consider a player with 
$\epsilon>0$. He cooperates with those of $W=0$ with probability $\epsilon>0$, and with those of $W=1$ cooperates for sure.
In the average, he cooperates with probability $(1+\epsilon)/2$, what is larger than 0.5. His reputation could increase, but
according to the curve $W(\epsilon)$ it is already equal to 1. This is a proof that indeed the curve $W(\epsilon)$ is stationary.
This result is consistent with the numerical results shown in Fig. 3, where the correlation between the reputation $W(\epsilon)$ 
and the altruism $\epsilon$ is shown. \\

\begin{figure}
\begin{center}
\includegraphics[scale=0.45,angle=-90]{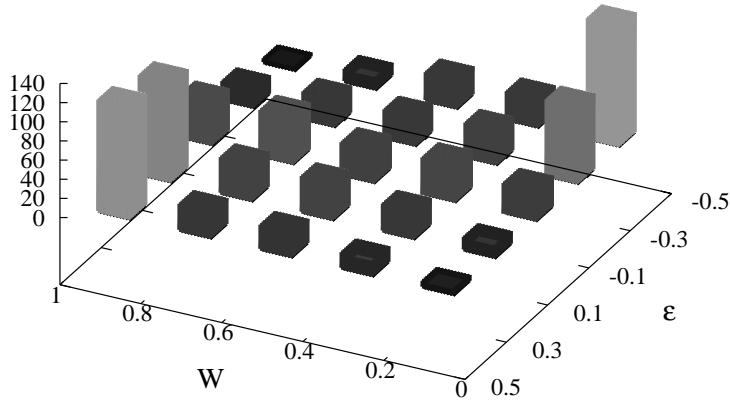}\\
\end{center}
\caption{Correlation between the reputation $W$ and the altruism $\epsilon$. The conditions of the calculation are the same as in Fig.1.
The number of players with $W$ and $\epsilon$ within given ranges is shown in the vertical axis.}
\end{figure}

\begin{figure}
\begin{center}
\includegraphics[scale=0.45,angle=-90]{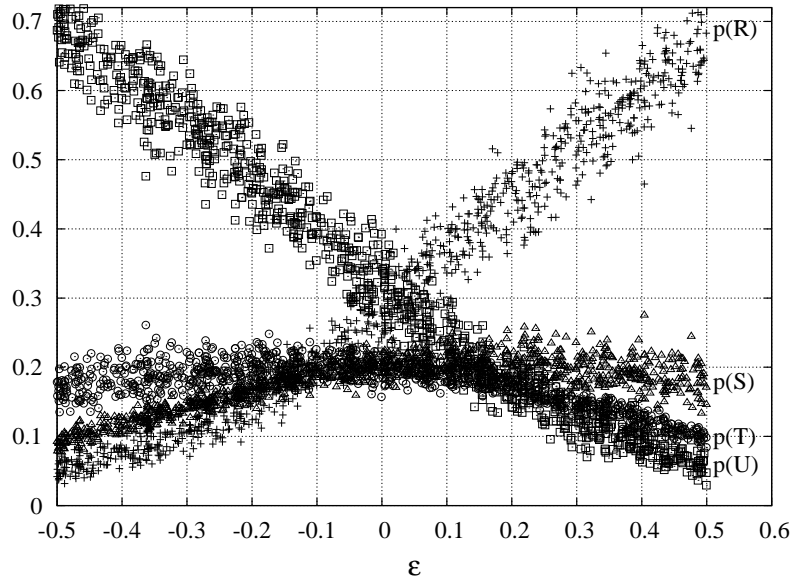}\\
\end{center}
\caption{Statistics of four different situations (R, S, T and U) against the altruism level $\epsilon$. The picture refers to the state 
after $5\times 10^5$ games played in the set of 1000 players. On the contrary to Fig. 1, only ten percent of links are present in 
the network. The games are played only along the links. 
}
\end{figure}

It is of interest to check also the role of the structure of the connections between players. In our model, best results 
are obtained if each player mets with each; this is the case of a fully connected graph. For the classical Erd\"os-Renyi 
network of 1000 players, the effect is somewhat disrupted for the mean degree $<k>$=100 (Fig.4).
This suggests that the mechanism is most effective when players have an access to the whole spectrum of co-players.

\section{Discussion}

Our results indicate that the cooperation can be established in the presence of two model ingredients. First is the altruism, 
understood as an enhanced attitude to cooperate. The enhancement means that this attitude excesses a purely rational evaluation 
of the probability that a co-player cooperates. The second ingredient is the reputation, what means that the information is 
scored about individual players. In this construction, the information on payoffs is omitted, what can mean that either
{\it i)} the players are not conscious about any payoff and they do not take it into account, or {\it ii)} the payoffs are known
and motivate players to cooperate or not. In both cases, our results remain valid.\\

Simple as it is, our model is not universal in its particular details. The method of calculating the probabilities
of cooperation and defection assures them to be zero or one for some values of the parameters $W$ and $\epsilon$. This means that 
some players always cooperate with each other, and some others always defect each other. A question appears, if our mechanism
is resistent with respect to some lack of information, when the reputation is unknown. Having this question in mind, we started
the simulatuion from the case where the reputation $W=0.5$ for all players. As described above, we have observed that the 
cooperation between some players appears. After some time, the cooperation is established within a group of players, more firmly
between those with larger level $\epsilon$ of altruism. We can conclude that although the reputation of this or the other player 
is reduced by a clash with a notorious defector, the mechanism persists in the group as a whole. This should be true even if the 
method of calculating the probabilities $P(i,j)$ is quantitatively different.\\

Finally, our results suggest that the positive feedback between the reputation and the altruism is particularly sensible to the contact
between players with possibly large levels of altruism. This is the necessary condition of an increase of their reputations, what
in turn enhances the probability of their cooperation and so on. In our model, each pair of connected players interacts with 
equal probability. Then, any dilution of the (social) network which disregards their cooperation, as this presented in Fig. 4, 
makes the feedback less efficient. However, in real social networks contact is maintained mostly between players who cooperate. 
As it was formulated by Mark Granovetter "the strength of a tie is
a (probably linear) combination of the amount of time, the emotional intensity, the intimacy (mutual confiding), and the reciprocal
services which characterize the tie." \cite{gran2}. For Granovetter, it was natural to consider rather positive than negative social 
links (see the footnote in \cite{gran2} to the above citation). Then, the departure from the fully connected network to a more 
realistic social structure should be done by an elimination of bonds between those who are not willing to cooperate. Provided that 
cooperators are available, isolation of defectors is less costful than punishment. If the social structure is taken into account 
in this proper way, our feedback is expected to be more strong than it can be deduced from our numerical results.\\

A comment should be added on the relation between the model described here and theories of cooperation, known in literature:
kin or group selection, direct and indirect reciprocity \cite{nosi}. Our point of view is that some overlaps exist between our feedback
mechanism and at least some of these theories. The concept of indirect reciprocity seems to be most close, because in our model 
the reputation is an attribute of an individual player, and the information about it is commonly accessible. On the other hand,
our feedback seems a convenient starting point to a group formation process. Similar approach to the Heider balance problem was
summarized in \cite{k3}. Finally, the kin selection theory could be also addressed within our model by an enhancement of initial values
of reputations within the same family or tribe. Among these possibilities, the group formation seems most attractive for further research.

\bigskip

{\bf Acknowledgements.} The research is partially supported within the FP7 project SOCIONICAL, No. 231288.

\end{document}